\documentclass[
aps,%
11pt,%
final,%
notitlepage,%
oneside,%
onecolumn,%
nobibnotes,%
nofootinbib,%
superscriptaddress,%
noshowpacs,%
centertags]%
{revtex4}

\begin{document}
\selectlanguage{english}

\title{Fundamental Plane of Groups and Clusters of Galaxies: Distances and  Peculiar Velocities of Superclusters of Galaxies on Small Scales}

\author{\firstname{F.~G.}~\surname{Kopylova}}
 \email{flera@sao.ru}
\affiliation{\saoname}

\author{\firstname{A.I.}~\surname{Kopylov}}
\affiliation{\saoname}

\begin{abstract}
This work is a continuation of the work of Kopylova and Kopylov
(2016) to build a fundamental plane (FP) of groups and clusters of
galaxies --- here a sample of galaxies systems is increased from 94
to 172 objects. We have studied the ratios between the basic characteristics
of groups and clusters of galaxies according to the archival data of SDSS,
2MASX and NED catalogs. Measured parameters ($\log L_K$, $\log R_e$ and
$\log \sigma$) of clusters of galaxies determine the fundamental plane
in the near infrared region:
$L_K \propto R_e^{0.77\pm0.09} \sigma^{1.44\pm0.12}$.
The form of the FP of groups/clusters is consistent with the FP of the
early --- types of galaxies (SDSS, $r$-band) determined in the same way.
Direct regression relative to the padius $\log r_e$ in the kpc gives a
projection of the FP --- 
$\log R_e=0.98(\pm0.06)\,\log \sigma-0.56(\pm0.04)<\log \langle I_e
\rangle+3.57(\pm0.07)$,
that can be used to determine the distances of the systems of galaxies.
The root-mean-square deviation of the FP zero-point is 0.07 which is equal
to the 16\% error in determining the distance of the group or cluster of
galaxies. For the first time, we measured the peculiar velocities  of the
superclusters of galaxies. The mean peculiar velocity of the 5 superclusters
of galaxies relative to the CMB is $+75\pm360$~km~s$^{-1}$.

\end{abstract}

\maketitle

\section{INTRODUCTION}

The large-scale structure of the Universe depends on cosmological parameters---cosmological density of matter, the growth rate of structures, and the Hubble expansion rate. Irregularities in the large-scale distribution of matter in the Universe cause peculiar motions of galaxies and galaxy clusters. Thus, the field of peculiar velocities of galaxies and galaxy clusters is a powerful tool in cosmological research.

The gravitationally bound systems---globular clusters of stars, galaxies, clusters of galaxies---obey the virial theorem in the form $M \propto R\sigma^2$, have similar relationships between parameters, for example, the fundamental plane. The fundamental plane is the empirical relationship between three observable characteristics of early-type galaxies: physical size ($R$), radial velocity dispersion along the line of sight ($\sigma$) and surface brightness. The FP is expressed through these quantities as $$\log R_e = a \log \sigma+b \log\langle I_e\rangle+c,$$ where $I_e$ is the average surface brightness within the effective radius $R_e$, $\sigma$ is the central  velocity dispersion, $c$ is the zero point (Djorgovski and Davis, 1987; Dressler et al., 1987; Jorgensen et al., 1996; Pahre et al., 1998; Bernardi et al., 2003; Saulder et al., 2013). Since $\log R_e$ and $\log\langle I_e\rangle$ are determined by the luminosity (or mass) of galaxies (the main parameter describing the galaxy)--- \linebreak $\langle I_e \rangle  = L/(2\pi R_e^2)$, then $\log L$, $\log \sigma$ and $\log R_e$ also form an FP--- $$\log L = a\log \sigma+b \log R_e+c.$$ The fundamental plane of early-type galaxies is often used to determine the peculiar velocities of galaxy systems (e.g., Jorgensen et al., 1996; Mohr and Wegner, 1997; Colless et al., 2001; Blakeslee et al., 2002;  Hudson et al., 2004; Batiste and Batuski, 2013;  O'Mill et al., 2015; Kopylova and Kopylov, 2017, 2021; Mutabazi, 2021).

For the first time, the FP of galaxy clusters themselves was constructed using photometric observational characteristics for a sample of 16 rich galaxy clusters  \mbox{($z<0.2$)} (Schaeffer et al., 1993). This work uses the luminosities $L$ in the $V$ filter and the effective radii $R_e$ of galaxy systems containing half the luminosity determined from the Vaucouleurs profile. As a result of calculating the regression dependence of $L$ on other parameters, the following type of FP was obtained: $L\propto R_e^{0.89\pm0.15} \sigma^{1.28\pm0.11}$. In the work of Adami et al. (1998), the studied parameters of twenty (ENACS sample) galaxy clusters were determined for different profiles: King, Hubble, NFW and Vaucouleurs. For the King profile that best fit the observed data, they obtained the FP \mbox{$L\propto R_e^{1.19\pm0.14} \sigma^{0.91\pm0.16}$}, for the Vaucouleurs profile (a better fit to the results of Schaeffer et al. (1993))---\mbox{$L\propto R_e^{0.61\pm0.28} \sigma^{0.95\pm0.32}$}.

For the WINGS sample of galaxy clusters \linebreak (D'Onofrio et al., 2013), surface brightness profiles were constructed in the $V$ filter, into which the Vaucouleurs profile is inscribed in the central region, and an exponential disk in the outer regions. The resulting FP has the following form (shown in the caption of the $y$ axis in Fig.~5 in D'Onofrio et~al. (2013)): 
\vspace{-2mm}
\begin{equation}
\begin{array}{rcl}
\log R_e &\! =\! &\!1.08(\pm0.16)\log\sigma-0.96(\pm0.13)\log \langle I_e\rangle \\[-4pt]
        &\! +\!&\!2.60(\pm0.47). \\[-10pt]
\end{array}
\end{equation}
A detailed study of the similarity of properties (parallelism) of early-type galaxies and galaxy clusters was carried out in the works of Chiosi et al. (2020), D'Onofrio et al. (2019, 2020).

The goal of this work is to construct a fundamental plane between the parameters of 172 groups and clusters of galaxies that have redshifts in the range 0.012 < $z$ < 0.10 and the radial velocity dispersions  200~km\,s$^{-1}$ < $\sigma$ < 1100~km\,s$^{-1}$. This work, with an increase in the number of objects, is a continuation of the work of Kopylova and Kopylov (2016). The characteristics we determine $L_K$, $\sigma$, $R_e$ (found from $N_{\rm tot}/2$) and the relationships between them (the fundamental plane) give us the opportunity to measure the relative distances of galaxy clusters in superclusters, the galaxy superclusters themselves, clarify the structural properties of galaxy clusters, the degree of their deviation from the relaxed state  evolutionary status). The sample includes 71~groups of galaxies with the radial velocity dispersion $\sigma \leq$ 400~km\,s$^{-1}$ and 101 galaxy clusters with $\sigma > 400$~km\,s$^{-1}$. Groups of galaxies are often defined as systems of galaxies with mass \mbox{$M<10^{14}$~$M_{\odot}$} and, accordingly, lower radial velocity dispersions \mbox{$\sigma < 400$~km\,s$^{-1}$}(e.g., Poggianti et al., 2008).

Previously, in the work of Kopylova and Kopylov (2016), we showed that when constructing the general FP of groups and clusters, it becomes necessary to search for an effective radius containing half of all galaxies within the halo radius ($R_{\rm sp}$). Because often the effective radii containing half the luminosity of galaxy groups ended up inside the central brightest galaxy.  Classically, in the system of standard photometric parameters of Vaucouleurs (de Vaucouleurs and Page, 1962), the effective radius of the galaxy $r_e$ is the radius of the circle within which half the total luminosity of the galaxy is emitted. The effective surface brightness is the surface brightness at a distance $r_e$ from the center of the galaxy. Kopylova and Kopylov (2016) also show that between the total luminosity $L_K$  and the number of galaxies $N$  (within a radius of $R_{200}$), defined up to the limit $M_K <- 21^{\rm m}$ for all systems there is a linear dependence on the luminosity function.

Groups and clusters of galaxies are located in the regions of the superclusters of galaxies Leo, Hercules, Ursa Major, Corona Borealis, Bootes, A\,1656/\,A1367. We also included in the sample rich X-ray clusters from other smaller superclusters of galaxies and fields. The dimensions of the dark halo of galaxy systems, the number of galaxies, luminosity and radius within which half of the galaxies are observed are determined from the integral distribution of the number of galaxies (according to the profile of galaxy systems) depending on the square of the radius from the center. The work was carried out by us using data from the  catalogs SDSS\footnote{\url{http://www.sdss.org}} (Sloan Digital Sky Survey Data Realease~7, 8 (Aihara et al., 2011)), 2MASS XSC (Two-Micron ALL-Sky Survey Extended Source Catalog  (Jarrett et al., 2000)) and NED\footnote{\url{http://nedwww.ipac.caltech.edu}} (NASA Extragalactic Database).

The article is organized as follows. The second section describes the sample of galaxy systems. In the third section, the parameters of galaxy systems necessary for constructing the FP are determined: first, the halo boundary $R_{\rm sp}$ is found, the number of galaxies within this radius is found, the effective radius $R_e$ is measured, within which half of the galaxies are observed, and the total luminosity $L_K$ is measured. In the fourth section, the general FP and FPs of members of galaxy superclusters are obtained, and FP relative to the long axis $\log R_e$, which are used to determine the distances of galaxy systems, are given. The fifth section describes the determination of distances and peculiar velocities of superclusters of galaxies. The conclusion lists the main results.

In our work we used the following cosmological parameters: $\Omega_m=0.3$, $\Omega_{\Lambda}=0.7$, $H_0=70$~km\,s$^{-1}$ Mpc$^{-1}$.

\section{DESCRIPTION OF THE SAMPLE}

Our sample consists of 172 groups and clusters of galaxies from the regions of superclusters of galaxies Leo ($N=12$), Hercules (Her, $N=27$), Ursa Major (UMa, $N=19$), Corona Borealis (CrB, \mbox{$N=8$}), Bootes (Boo, $N=13$), as well as other smaller superclusters and fields ($N=25$ and $N=20$, respectively), groups of galaxies from the region of the supercluster A\,1656/\,A1367 ($N=48$). A sample of galaxy clusters in superclusters was compiled to measure the peculiar velocities of galaxy clusters within them along the fundamental plane of early-type galaxies (Kopylova and Kopylov, 2007, 2014, 2017, 2021). In addition, we investigated the relationship between dynamic mass within the virialized radius $R_{200}$ and infrared (IR) luminosity in the $K_s$ filter (hereinafter $K$) (Kopylova and Kopylov, 2009, 2011, 2013, 2015, 2022 ). All galaxy clusters have detected X-ray emission, except for 21 galaxy groups. The sample of galaxy systems covers the maximum range of system radial velocity dispersions---from 200~km\,s$^{-1}$ to 1100~km\,s$^{-1}$---in the local Universe (0.01 < $z$ < 0.09).

Processing for all sample objects was carried out in the same way. The corresponding empirical cluster radius $R_{200}$ is predicted by the radial velocity dispersion of galaxies and can be estimated by the formula \mbox{$R_{200} = \sqrt {3} \sigma /(10H(z))$}~Mpc (Carlberg et al., 1997). Then, assuming that the cluster is virialized within this radius, one can find the mass \mbox{$M_{200} = 3G^{-1}R_{200}\,\sigma^{2}$}, where $\sigma$ is the one-dimensional dispersion of radial velocities of galaxies located within the radius $R_{200}$, $G$ is the gravitational constant. For galaxy systems, we measured heliocentric redshifts, radial velocity dispersions with cosmological correction $(1+z)^{-1}$, radii $R_{200}$, \mbox{$K$-luminosities} $L_{K,200}$ \mbox{($M_K < -21^{\rm m}$)}, dynamic masses $M_{200}$ and other parameters of galaxy systems within the radius $R_{200}$, which are given in the indicated works.

\section{DETERMINATION OF THE PARAMETERS OF THE FUNDAMENTAL PLANE OF GROUPS AND CLUSTERS OF GALAXIES}
\subsection{Halo radius---splashback radius $R_{\rm sp}$, effective radius $R_e$}

Similar to the fundamental plane of early-type galaxies, the fundamental plane of galaxy groups and clusters is the empirical relationship between the radial velocity dispersion of galaxies within a chosen radius, the physical effective radius, and the mean surface brightness within the effective radius. In Kopylova and Kopylov (2016), we showed that in galaxy clusters it is possible to find an effective radius containing half of the galaxy haloes to construct a common fundamental plane for both poor groups and rich galaxy clusters.

To do this, for groups/clusters of galaxies, we determine the splashback radius $R_{\rm sp}$, according to the methodology outlined in Kopylova and Kopylov (2016), and find the number of galaxies within it, corrected for the background (determined by the slope of the straight line). We then determine the radius containing half the galaxies and calculate the luminosity of these galaxies. By doubling it, we obtain the total luminosity of the group or cluster. For most rich clusters of galaxies (but not groups) with \mbox{$\sigma > 400$~km\,s$^{-1}$} the effective radius can be determined in the usual way---as the radius within which half the luminosity of the system is contained. We have shown that effective radii measured by the number of galaxies are on average 17\% larger than radii measured by luminosity (Kopylova and Kopylov, 2016).

To find the halo radius $R_{\rm sp}$, it is important for us to identify the immediate neighborhoods of galaxy clusters. For this purpose, we use a set of figures that characterize in detail the structure and kinematics of galaxy clusters. As an example, in Fig.~1 we show the one for A\,1668. Information on the main physical parameters of the cluster can be found in Kopylova and Kopylov (2022). The panels of Fig.~1 present:

\begin{list}{}{
\setlength\leftmargin{8mm} \setlength\topsep{2mm}
\setlength\parsep{0mm} \setlength\itemsep{2mm} }
    \item[(a)] deviation of the radial velocities of galaxies---members of the cluster, and galaxies assigned to the background, from the average radial velocity of the cluster or group, depending on the square of the radius (distance from the center of the cluster);
    \item[(b)] location of galaxies in the plane of the sky in equatorial coordinates;
    \item[(c)] integrated distribution of the number of all galaxies depending on the square of the radius;
    \item[(d)] histogram of the radial velocity distribution of all galaxies within the radius $R_{200}$.
\end{list}

\begin{figure*}[ht!]
\setcaptionmargin{5mm} \onelinecaptionstrue \captionstyle{normal}
\includegraphics[scale=0.57, angle=-90]{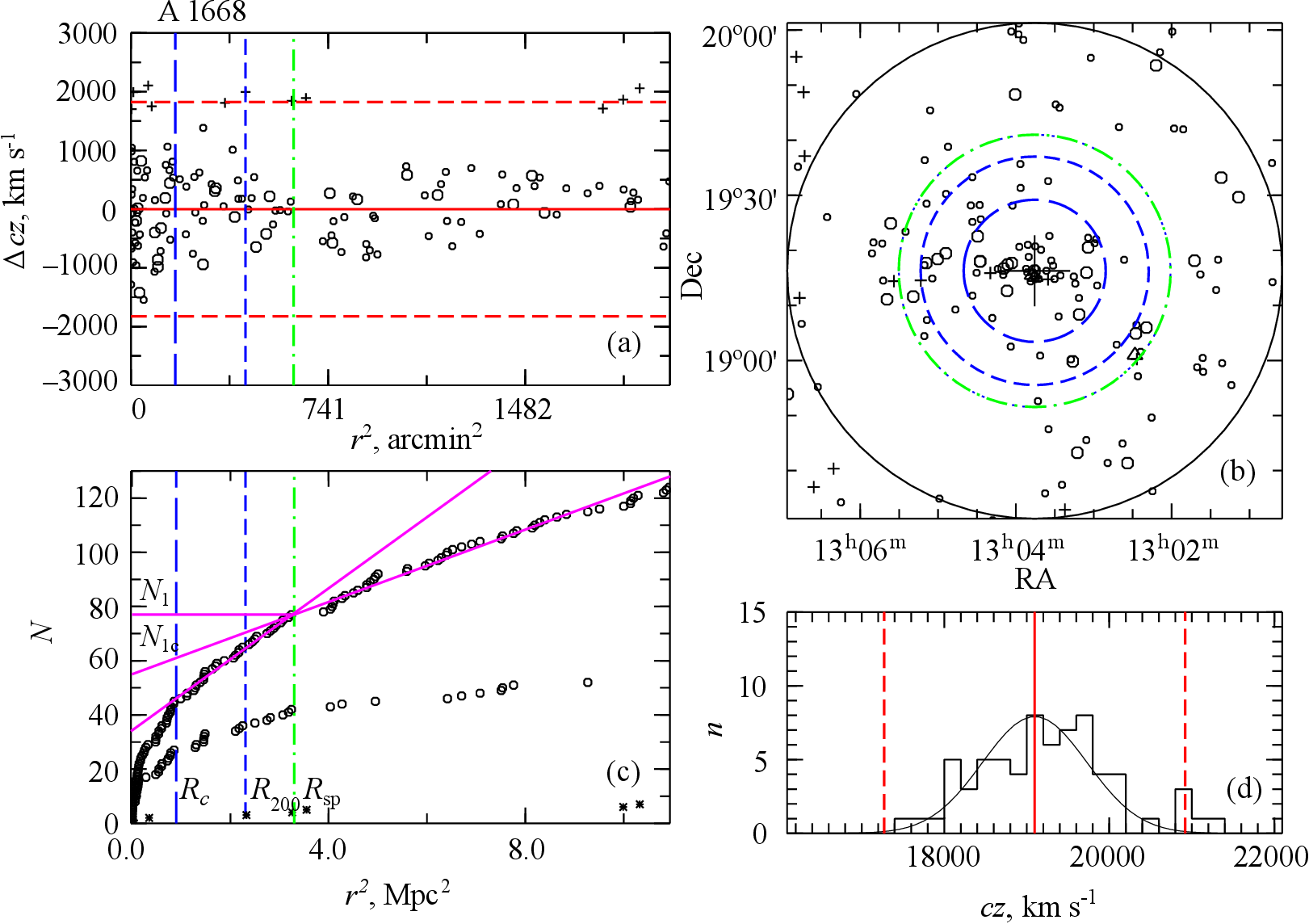}
\caption{Distribution of galaxies in the A\,1668 cluster. Panel (a) shows the deviation of the radial velocities of galaxies from the average radial velocity of the cluster determined from galaxies within the radius $R_{200}$. Horizontal dashed lines correspond to deviations $\pm2.7\sigma$, vertical lines show: radii $R_{200}$ and $R_c$---short and long dashes, respectively, dash-dotted line---radius $R_{\rm sp}$. Larger circles indicate galaxies  brighter than $M_K^*+1 = -24^{\rm m}$, straight crosses---background galaxies, oblique crosses---foreground galaxies. The abscissa shows the  squared distance from the center of the cluster in arcminutes. Panel (b) presents the distribution on the sky in the equatorial coordinate system of the galaxies that are presented in panel (a) (the designations are the same). Circles highlight areas of radii $R_{200}$ (short dashes),  $R_c$ (long dashes), $R_{\rm sp}$ (dash-dotted line). The study area is  limited by a circle with a radius of $3.5\,R_{200}$ (solid line). The  center of the cluster is marked with a large cross. Panel (c) shows  the integrated distribution of the total number of galaxies (upper curve) as a function of the squared distance from the cluster center (in squared megaparsecs). The lower curve is the distribution of early-type galaxies brighter than $M_K=-21\,.\!\!^{\rm m}5$. The  circles correspond to the galaxies indicated by the circles in panel (a), the asterisks---to the background galaxies. Solid lines characterize the course of the distribution of galaxies behind and in front of the radius $R_{\rm sp}$. $N_1$ and $N_{\rm 1c}$ stand for the number of galaxies Panel (d) shows the radial velocity distribution of all galaxies within the $R_{200}$ radius (the solid line for cluster members shows the Gaussian corresponding to $\sigma$ of the cluster). The solid vertical line indicates the average radial velocity of the cluster, the dashed lines correspond to deviations of $\pm2.7\sigma$.}
\end{figure*}

Of particular interest is Fig.~1c, which shows the integrated distribution of the number of galaxies as a function of the squared distance from the center of the cluster A\,1668, coinciding with the brightest galaxy. The indices $c$, $200$ and sp in Fig.~1 indicate the radii in megaparsecs of the cluster regions we have identified (the core, the virialized region and the dark matter halo). As modeling shows, the radius $R_{\rm sp}$ we found is the actual boundary of the dark matter halo (Adhikari et al., 2014; Diemer and Kravtsov, 2014).

Thus, analysis of Fig.~1 allows us to conclude that the main part of the galaxies in the cluster is located in the region limited by the radius $R_{\rm sp}$. Our task is to select this region, find the number of galaxies in it taking into account the background, calculate their IR luminosity also taking into account the background and estimate the radius of the cluster as the radius containing half of the galaxies up to our chosen limit \mbox{$M_K = -21^{\rm m}$}. As shown in Fig.~1c, within the radius $R_{\rm sp}$ there are $N_1$ galaxies without taking into account the background, and taking into account the background---$N_{1c}$ galaxies that we found graphically by the inclination in the distribution of galaxies outside the cluster halo. $N_1$ and $N_{1c}$ are related by the equation $N_{1c} = N_1-\pi  R_{\rm sp}^2 \ \Sigma$, whence the density of galaxies (in the region $R_{\rm sp}$) is equal to  \mbox{$\Sigma = (N_1-N_{1c})/(\pi  R_{\rm sp}^2)$}. Using the resulting galaxies, we found the radius at which $N_{1c}/2$ galaxies are observed. In Fig.~2, using the example of a cluster close to A\,1656 ($z$~=~0.024), the dependence of the radius $R_{\rm sp}$ on the absolute size of galaxies in the $K$~filter is shown. It can be noted that in the cluster the radius $R_{\rm sp}$ does not depend on the limiting sample size, except, perhaps, for a subsample of the brightest galaxies with $M_K<-23\,.\!\!^{\rm m}35$. Our sample limit practically corresponds to the most distant galaxy systems A\,2142 ($M_K<-23\,.\!\!^{\rm m}35$) and A\,2244 ($M_K<-23\,.\!\!^{\rm m}5$).

\begin{figure*}[ht!]
\setcaptionmargin{3.5mm} \onelinecaptionstrue \captionstyle{normal} \vspace{2mm}
\includegraphics[scale=0.63,angle=0]{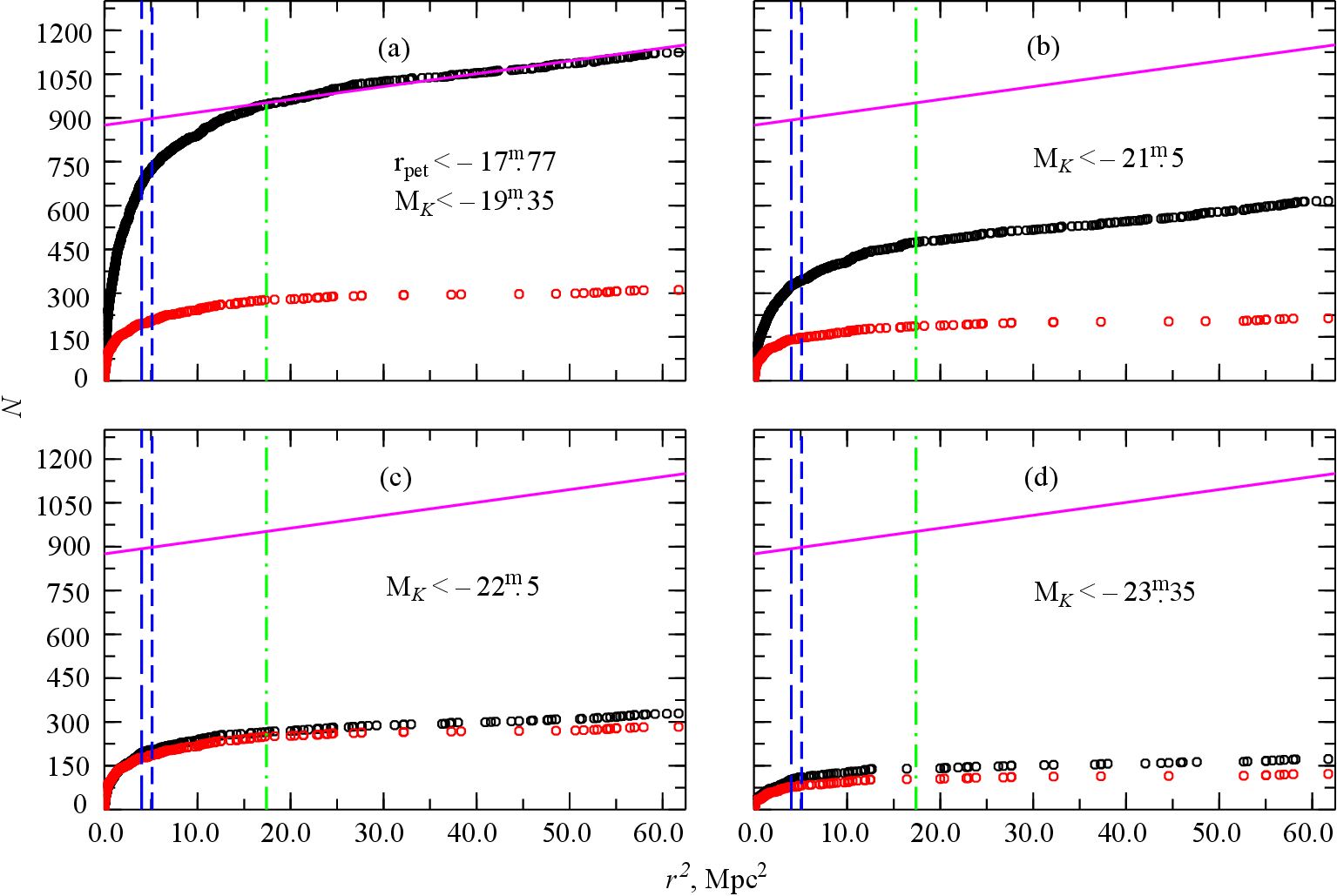}
\caption{Integrated distribution of the number of galaxies in the A\,1656 cluster depending on the squared distance from the center ($R<3.5R_{200}$) in megaparsecs (upper curve). The lower curve corresponds to early-type galaxies. Vertical lines show: short and blue dashes---radii $R_{200}$ and $R_c$, respectively, green dash-dotted line---radius $R_{\rm sp}$. Panel (a) shows all  galaxies extracted from the SDSS catalog ($r_{\rm pet}<-17\,.\!\!^{\rm m}77$, $M_K<-19\,.\!\!^{\rm m}35$). Panels (b)--(d) show galaxies corresponding to luminosities: $M_K<-21\,.\!\!^{\rm m}5$, $M_K<-22\,.\!\!^{\rm m}5$ and $M_K<-23\,.\!\!^{\rm m}35$, respectively.}
\end{figure*}

The next step in constructing the FP is to determine the total luminosity of galaxy systems to the same limit \mbox{$M_K = -21^{\rm m}$}. In groups and clusters of galaxies, the most massive and bright galaxies are early-type galaxies, that is, bright galaxies dominate the total luminosity of clusters, but faint galaxies make the main contribution to the total number of galaxies.

\subsection{Total $K$-luminosity}

The infrared emission from stars is not significantly affected by either a burst of star formation in the galaxy or dust, since the central regions of galaxy clusters contain mainly early-type galaxies with old stellar populations. Therefore, IR emission more accurately tracks the mass of the stellar population of galaxy clusters and is often used for this purpose (e.g., Lin et al., 2004; Ramella et al., 2004). To determine the total luminosity of galaxy clusters in the IR region, we used the photometric results presented in the  final version of the 2MASS\footnote{\url{http://www.ipac.caltech.edu/2mass/releases/allsky/}} catalog for extended objects (XSC, Jarrett et al., 2000). Since approximately a third of the galaxies of groups and clusters in our sample, detected spectrally in the SDSS catalog, did not have XSC measurements, to determine \mbox{$K$-values} we applied the method described by Obri\'{c} et al. (2006). For 99\,000 SDSS\,(DR1) galaxies, the value of $K_{\rm SDSS}$ was determined from the color index ($u-r$). The following scheme was used: \mbox{$K_{\rm SDSS} = r_{\rm pet}-(r-K)$}, where $r_{\rm pet}$ is the Petrosyan magnitude of the galaxy in the $r$ filter, and $r-K$ is determined by the formula
\vspace{-2mm}
\begin{equation}
\begin{array}{rcl}
r-K& = &1.115+0.94(u-r)-0.165(u-r)^2 \\[-4pt]
        & + &0.00851(u-r)^3+4.92z-9.1z^2,
\end{array}
\end{equation}
where $z$ is the redshift of the galaxy. 

In addition, corrections were introduced into the $r-K$ color, which are equal to $0.496-0.154R^z_{50}$ for early-type galaxies ($u-r>2.22$) and  $0.107-0.045R^z_{50}$ for late-type galaxies ($u-r \leq2.22$), where $R^z_{50}$ is the radius  bounding the region that emits $50\%$ of the Petrosyan flux in the $z$ filter. According to the method adopted in Graham et al. (2005), we adjusted the Petrossian magnitudes to the full magnitude of the galaxy using the formula:

\begin{equation}
r_{\rm tot} = r_{\rm pet}-5.1\times10^{-4} \exp((R^r_{90}/R^r_{50})^{1.451}),
\end{equation}
where $R^r_{90}$ and $R^r_{50}$ are the radii of the regions containing 90\% and 50\% of the Petrosian flux in the $r$ filter. 

According to our work (Kopylova and Kopylov, 2009), the difference between the calculated $K_{\rm SDSS}$ and the total $K_{\rm XSC}$ magnitudes obtained from isophote magnitudes corresponding to the surface brightness $\mu_K= 20$~mag\,arcsec$^{-2}$, amounted to $0\,.\!\!^{\rm m}12\pm0\,.\!\!^{\rm m}02$. When calculating $K_{\rm SDSS}$ we took this correction into account. In addition, the average error in the galaxy isophote magnitude from the catalog of extended objects for the clusters under study is $0\,.\!\!^{\rm m}1$.

The completeness of our sample is determined by the completeness of the spectral data of the SDSS catalogue. Subject to the conditions $r_{\rm pet}<17\,.\!\!^{\rm m}77$ and \mbox{$\mu_r~< 24.5$~mag arcsec${^{-2}}$} (Petrosyan magnitude of the galaxy in the $r$ filter, corrected for extinction in the Galaxy, and Petrosyan mean surface brightness corresponding to the effective radius) the completeness of SDSS data is estimated to be 99\% Strauss et al. (2002), and for bright galaxies---95\%. We added from NED the missing SDSS measurements of the radial velocities of bright galaxies (from 1 to 5 for different clusters).

The luminosity of the galaxy system $L_{200,K}$ is equal to the sum of the luminosities of the galaxies in the $K$ filter up to a fixed absolute value. We took the value $-21^{\rm m}$ as this limit. First, we converted the observed magnitudes of the galaxies into absolute magnitudes using the formula:
\begin{equation}
M_K = K-25-5 \log(D_{l}/{\textrm {1\,Mpc}})-A_K-K(z),
\end{equation}
where $D_l$ is the distance to the galaxy for calculating the luminosity, $A_K$ is the extinction in the Galaxy, $K(z)$ = $-6\log(1+z)$ ($K$-correction, according to Kochanek et al. (2001).

We did not take into account the correction for the evolution of luminosity, since the range of changes in redshifts in our sample is small. The extinction in the Galaxy that we obtained from NED for the studied galaxy clusters is less than $0\,.\!\!^{\rm m}01$. The 2MASS (XSC) catalog is not a deep survey (the photometric limit with completeness above 90\% is $13\,.\!\!^{\rm m}5$ in the $K$ filter (Jarrett et al., 2000). We supplemented this catalog with galaxies from the SDSS catalog under the condition \mbox{$r_{\rm pet}<17\,.\!\!^{\rm m}77$}. Taking into account that the color ($r-K$) of early-type galaxies, which constitute the majority within $R_{200}$, is on average $2\,.\!\!^{\rm m}8$, the limit of our sample of galaxies with individual estimates of magnitudes in the K band is $\sim15^{\rm m}$.

Determining the total luminosity of a galaxy cluster using the luminosity function (LF) within a selected radius consists of two steps, described in detail in Kopylova and Kopylov (2009): first, the LF is normalized to the observed number of galaxies, then extrapolated into the region of faint magnitudes to a selected limit. In this case, the parameters of the Schechter function ($M_K^*$---characteristic value and $\alpha$---slope) or fields are usually used, or they are determined for the composite LF of the sample under study.

We applied the second method and found the parameters of the Schechter function of the composite LF for part of our sample, the superclusters of galaxies Leo and Hercules (Kopylova and Kopylov, 2013). For this purpose, galaxy counts were made for each cluster in intervals equal to $0\,.\!\!^{\rm m}5$. Then, composite luminosity functions were constructed for the virialized regions of the galaxy clusters of the Hercules and Leo superclusters, using the method described in Colless (1989). The resulting composite LFs were approximated by the nonlinear least squares method with the Schechter function in the interval [$-26\,.\!\!^{\rm m}0$,$-21\,.\!\!^{\rm m}5$] (Schechter, 1976) with the parameters \mbox{$M^*_K=-24\,.\!\!^{\rm m}99\pm0\,.\!\!^{\rm m}27$}, \mbox{$\alpha =-1.23\pm0.08$} for Her and \mbox{$M^*_K=-24\,.\!\!^{\rm m}90\pm0\,.\!\!^{\rm m}30$}, \mbox{$\alpha =-1.14\pm0.11$} for Leo (Kopylova and Kopylov, 2013). On   \linebreak average for both superclusters we obtained \linebreak \mbox{$M^*_K=-24\,.\!\!^{\rm m}97\pm0\,.\!\!^{\rm m}30$}, $\alpha =-1.19\pm0.10$, and we used these parameters to find the total luminosity of galaxy clusters. Since our sample contains distant clusters of galaxies, we found the total luminosity for clusters with \mbox{$z > 0.035$} as the sum of the luminosities of all observed galaxies, and to it we added the addition (up to $M_K=-21^{\rm m}$) obtained from the composite luminosity function of galaxy clusters using the method described in Kopylova and Kopylov (2013).

Thus, for galaxies contained within the found effective radius of systems (the radius within which half of the galaxies found within $R_{\rm sp}$ are contained), we measured the total $K$-luminosity corrected by the stacked LF to the limit $M_K =-21^{\rm m}$ and multiplied by 2 to find the total luminosity of galaxy clusters, $\log L_K$.

\section{FUNDAMENTAL PLANE OF GROUPS, CLUSTERS AND SUPERCLUSTERS OF GALAXIES}

In order to be able to compare the results obtained for constructing cluster FPs with the results of other authors (for example, Schaeffer et al., 1993), we took $\log L_K$, $\log R_e$, $\log \sigma$ and found the regression relation relative to $\log L_K$. In our study, $\log L_K$ is the total luminosity of groups and clusters of galaxies in the $K$ filter ($M_K=-21^{\rm m}$) in solar luminosities $L_{\odot}$, $\log R_e$ is the effective radius of systems in kpc, within which half of the total number of galaxies ($N/2$) found from the previously described profile is contained. It is known that the surface brightness in this case is measured in luminosities $L_{\odot}$ from an area of 1~pc$^2$ or kpc$^2$.

In the luminosity of galaxy groups/clusters, a correction for the evolution of luminosity in the $K$ filter was introduced equal to $0.8\,z$ (Bell et al., 2003). Unlike the FP of early-type galaxies, where the central radial velocity dispersion of stars is considered, for clusters and groups of galaxies it is better to take the radial velocity dispersion of galaxies within the virialized region (in our case we took $\sigma_{200}$ within $R_{200}$) as the most accurate, since the accuracy depends on the number of galaxies used.

Figure~3 shows the FP of our entire sample of 172 groups and clusters of galaxies, and individual members of each galaxy supercluster are highlighted with large circles.  Red circles correspond to groups of galaxies, blue ones---clusters of galaxies. The line shows the regression relation obtained from galaxy clusters with \mbox{$\sigma > 400$~km\,s$^{-1}$}. It looks like:

\begin{equation}\vspace{-1mm}
\label{1}
\begin{array}{rcl}
\log L_K&\! =\!& 0.77(\pm0.09) \log R_e+1.41(\pm0.12) \log \sigma\\[-2pt]
&\!+\!&6.69(\pm0.11).
\end{array}
\end{equation}

Considering that the average surface brightness within the effective radius, measured in kiloparsecs,
$\langle I_e\rangle$ = $L_K/(2 \pi R_e^2)$, we obtain
\vspace{-0mm} 
\begin{equation}
\label{2}
\begin{array}{rcl}
\log R_e &\!=\!&\! 1.14(\pm0.10) \log \sigma-0.81(\pm0.02) \log \langle I_e \rangle \\[-4pt]
&\!+\!&\!4.78(\pm0.09).
\end{array}
\end{equation}

In the expanding Universe $I \propto (1+z)^{-4}$, where $z$ is the redshift of the object. If we make this correction to the surface brightness and do a regression on $\log R_e$, we get the following regression relation:
\vspace{-2mm}
\begin{equation}
\begin{array}{rcl}
\log R_e &\!=\!&\! 0.98(\pm0.06) \log \sigma-0.56(\pm0.04) \langle \log I_e\rangle\\[-4pt] &\!+\!&\!3.57(\pm0.07).
\end{array}
\end{equation}

\begin{figure*}[]
\setcaptionmargin{3.5mm} \onelinecaptionstrue \captionstyle{normal}
\includegraphics[scale=0.365,angle=-90]{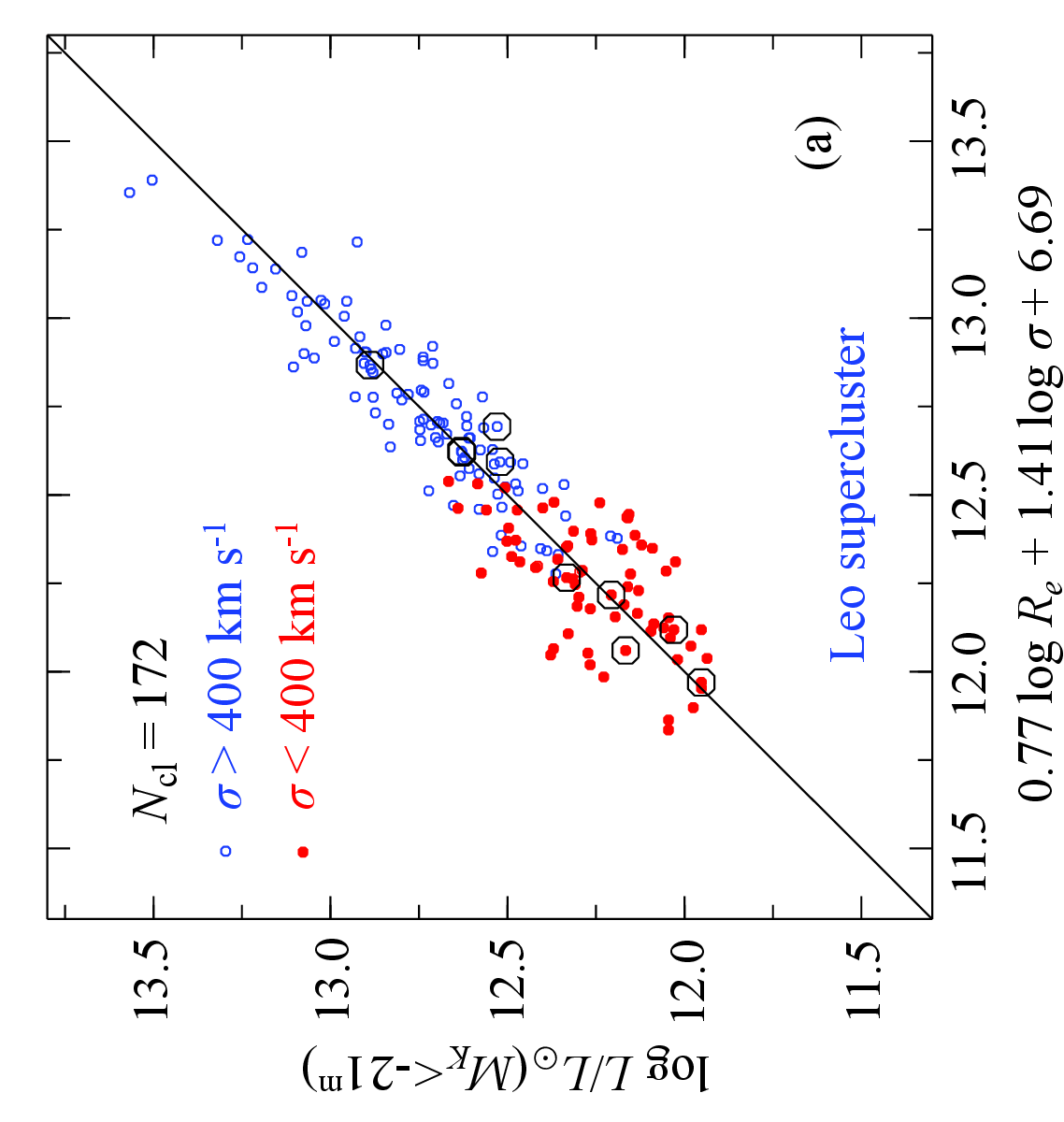}
\includegraphics[scale=0.365,angle=-90]{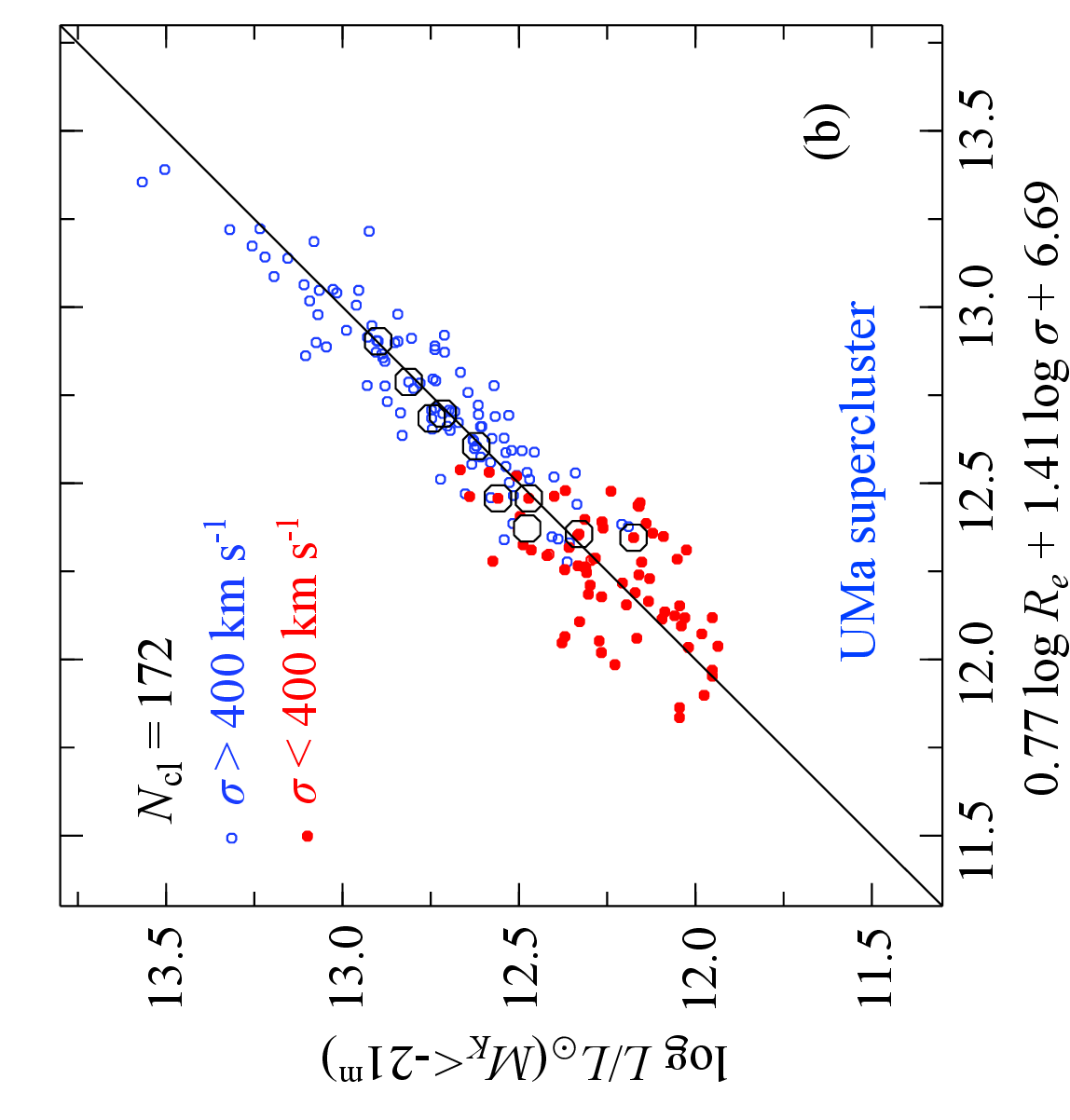}
\includegraphics[scale=0.365,angle=-90]{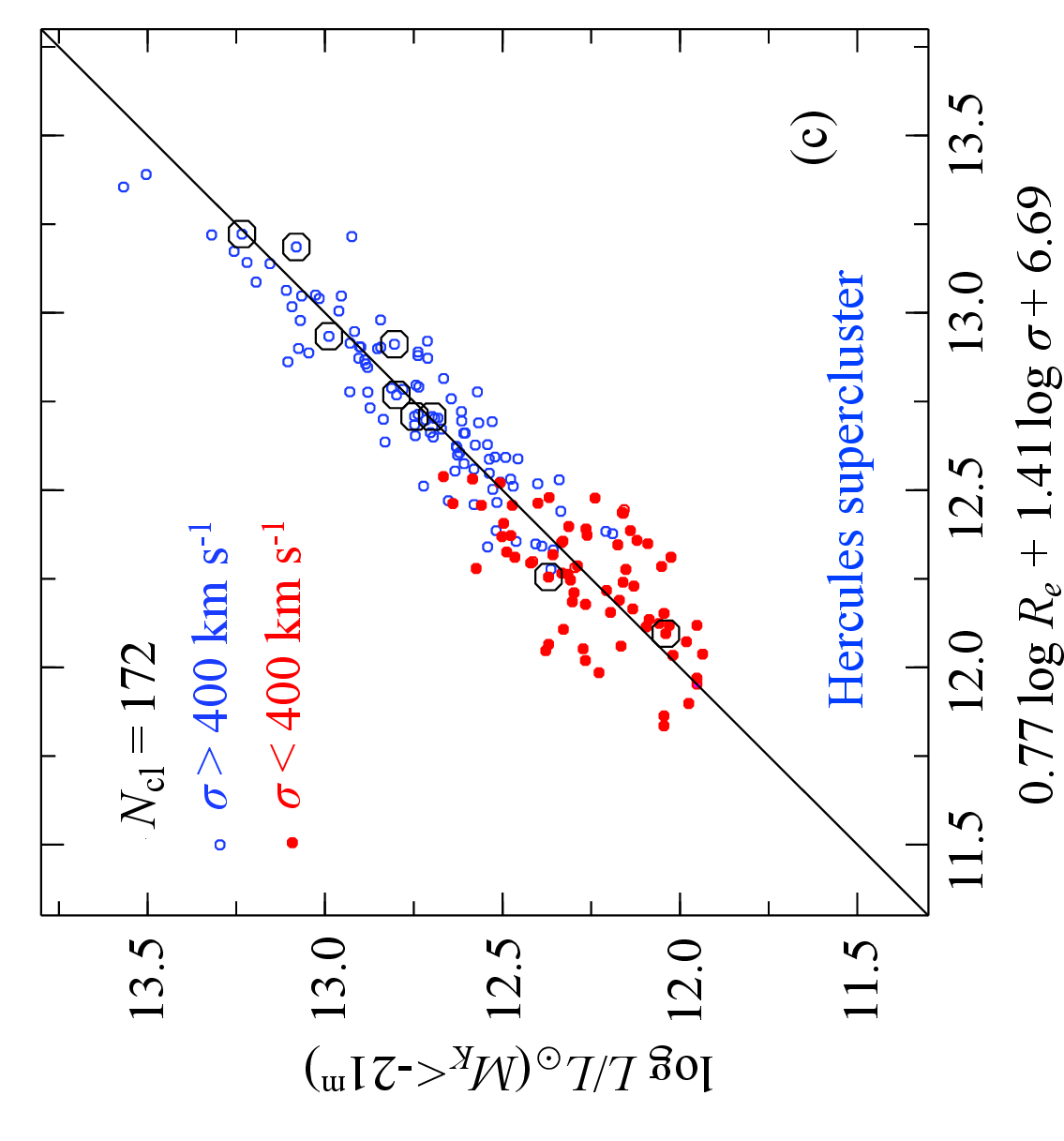}
\includegraphics[scale=0.365,angle=-90]{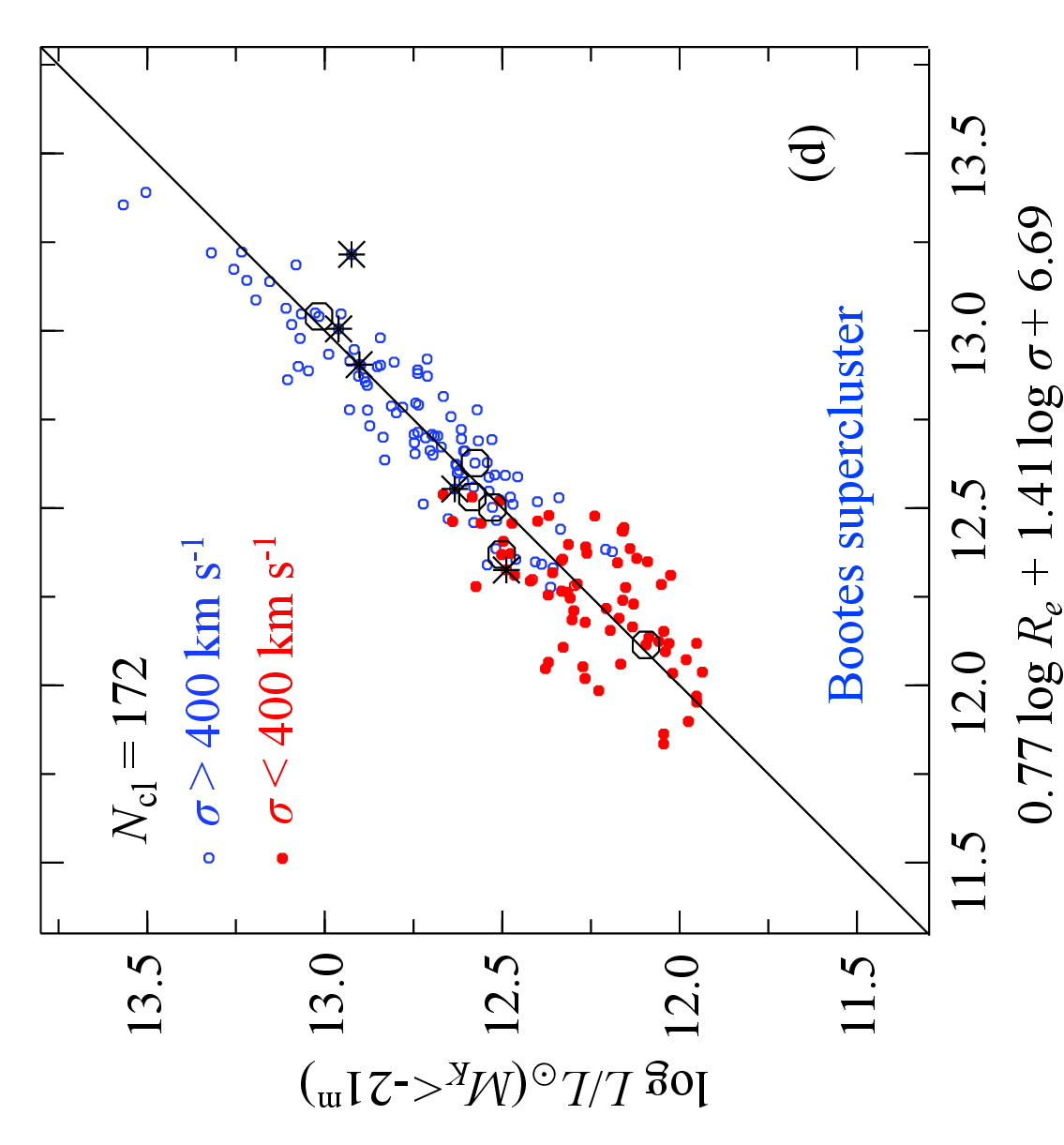}
\includegraphics[scale=0.365,angle=-90]{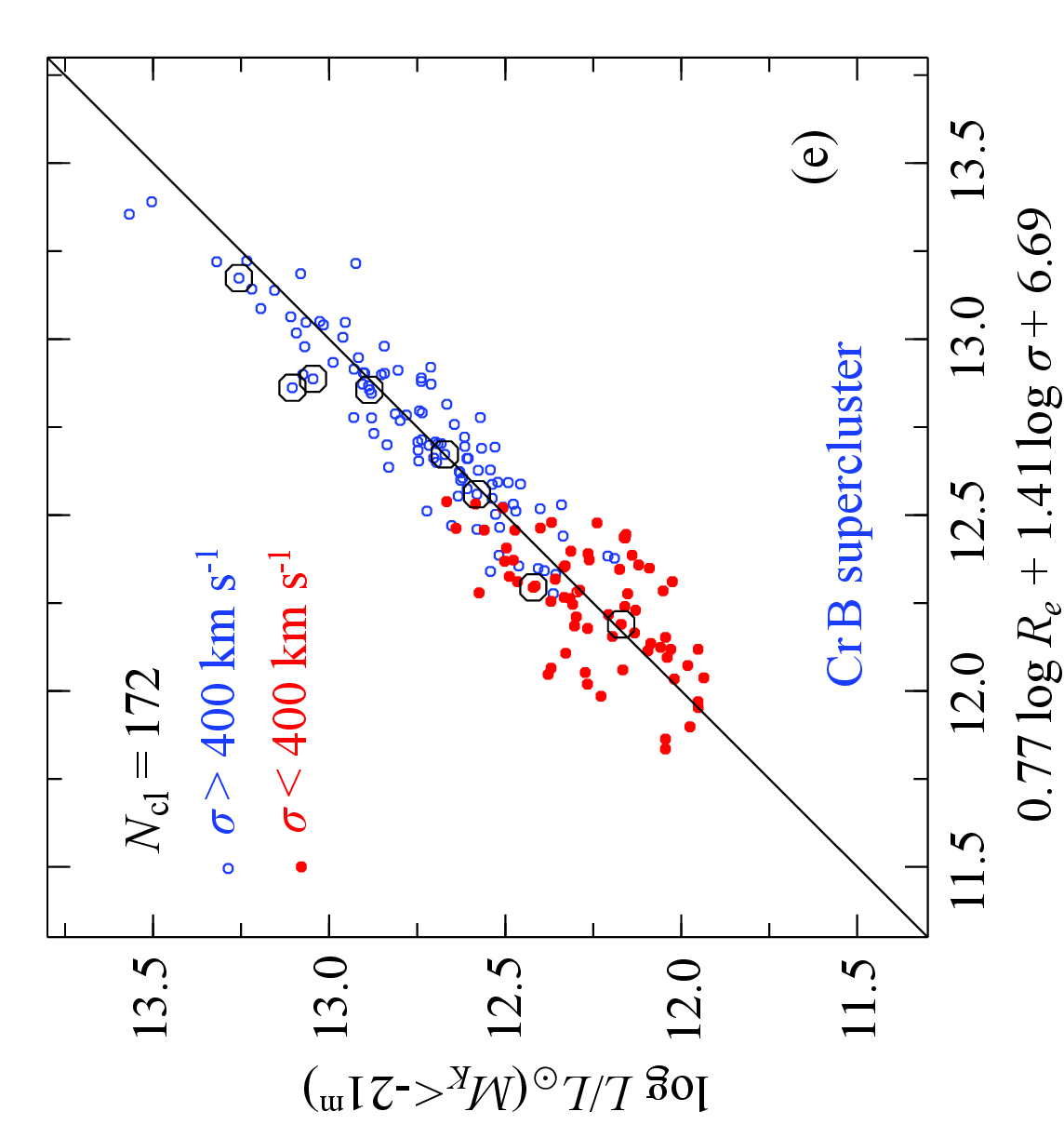}
\caption{Fundamental plane of 172 groups and clusters of galaxies in the
IR region. Members of the superclusters of galaxies Leo, Ursa Major,
Hercules, Bootes, Corona Borealis are circled and shown in panels (a)--(e),
respectively. In the Bootes system, crossed out crosses show clusters of
galaxies with $z = 0.076$, and empty circles---with $z = 0.064$. The line
corresponds to the regression relation
$L_K \propto R_e^{0.77\pm0.09} \sigma^{1.41\pm0.12}$.}
\end{figure*}

\begin{figure*}[]
\setcaptionmargin{3.5mm} \onelinecaptionstrue \captionstyle{normal}
\includegraphics[scale=0.365,angle=-90]{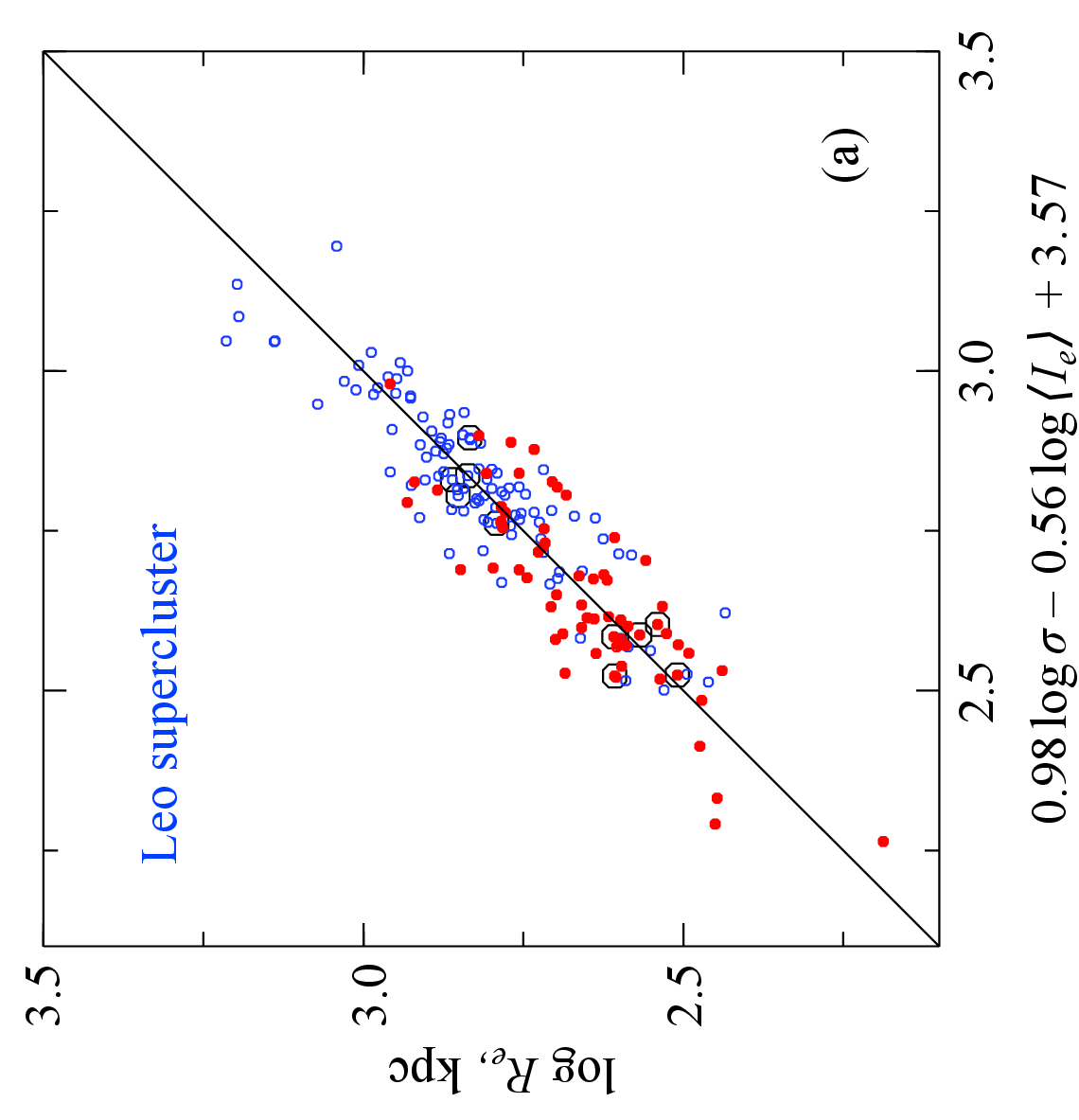}
\includegraphics[scale=0.365,angle=-90]{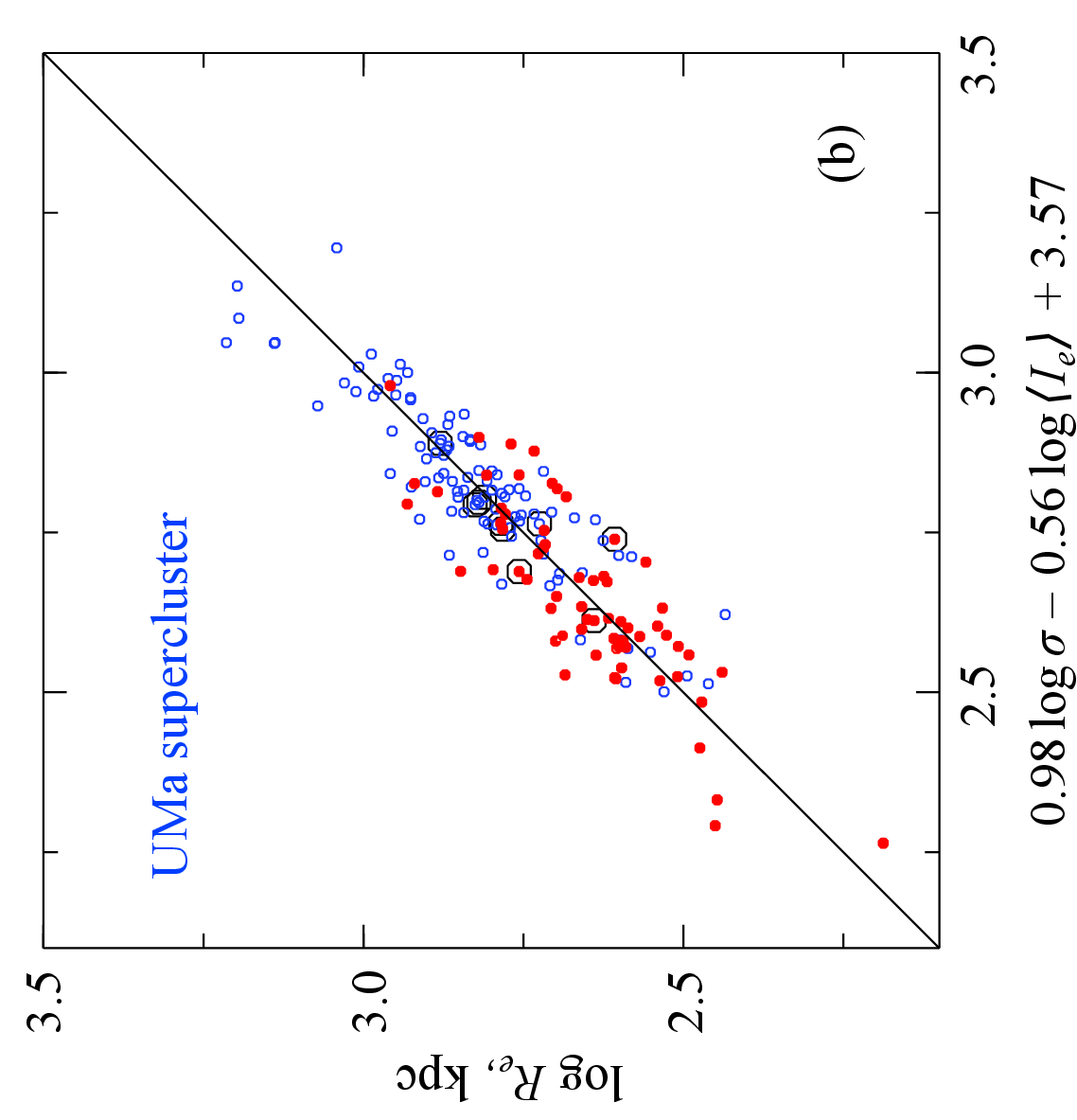}
\includegraphics[scale=0.365,angle=-90]{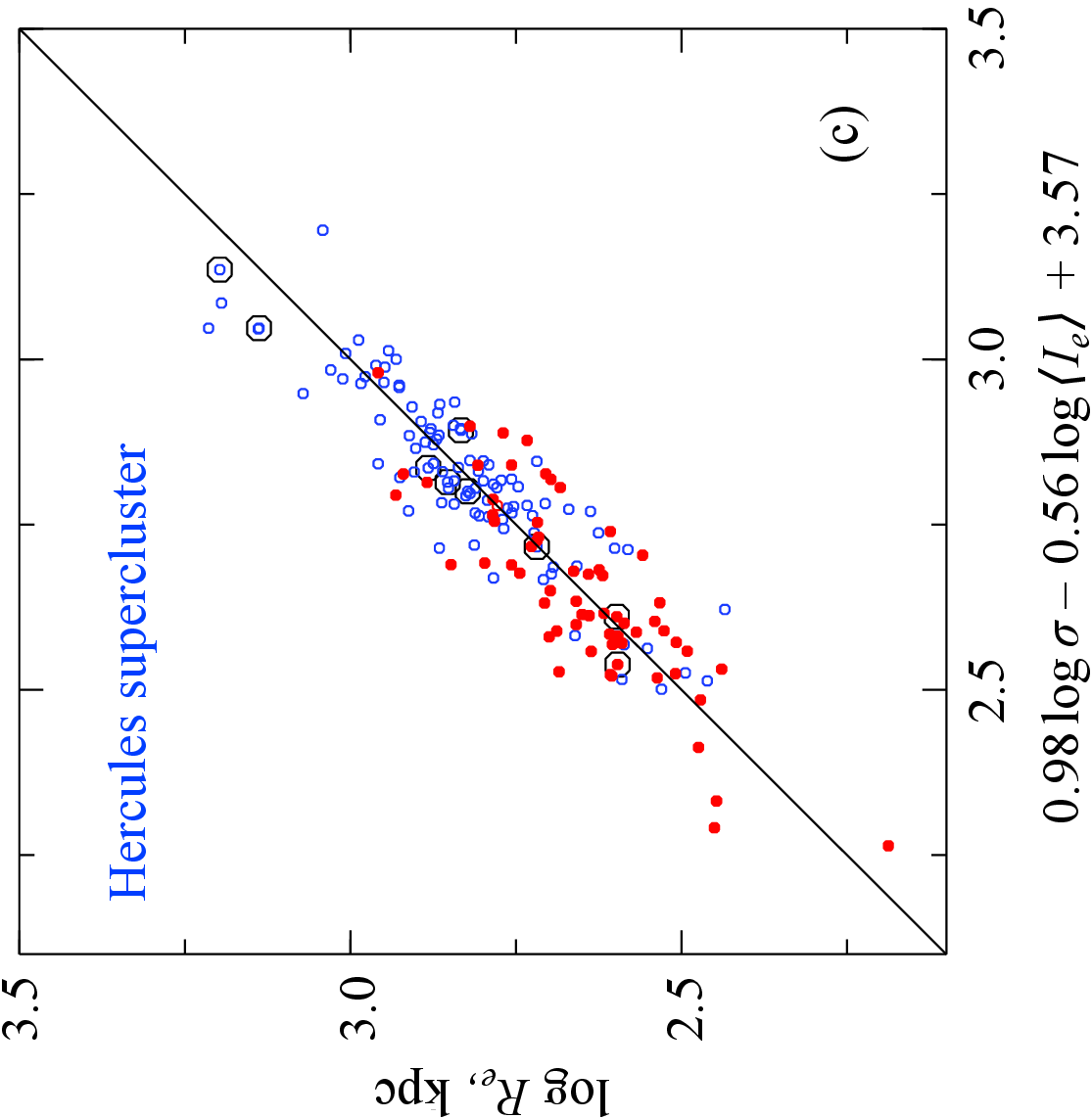}
\includegraphics[scale=0.365,angle=-90]{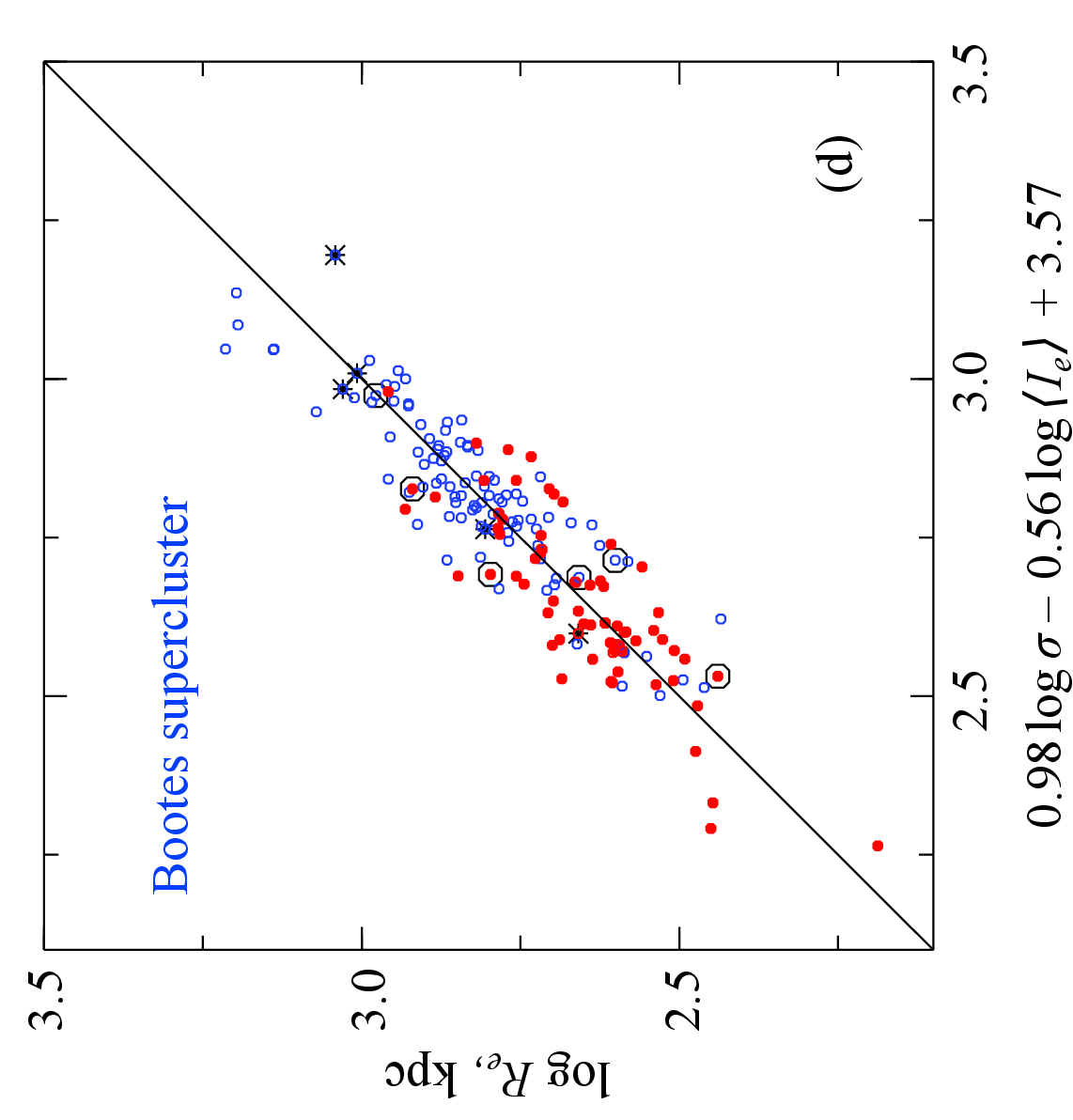}
\includegraphics[scale=0.365,angle=-90]{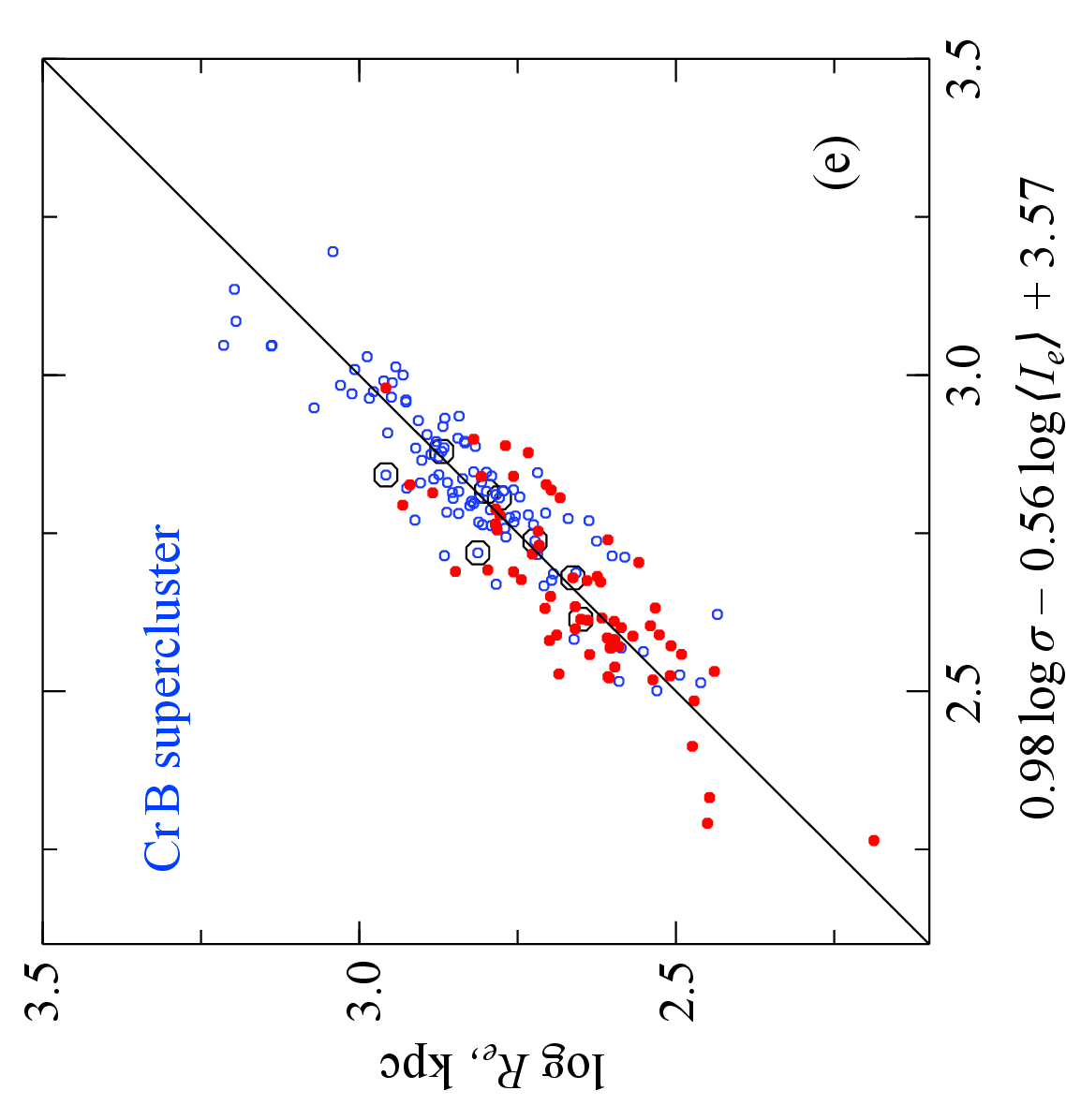}
\caption{Fundamental plane of 172~groups and clusters of galaxies in the
IR region along the long axis, $\log R_e$. Members of the superclusters of
galaxies Leo, Ursa Major, Hercules, Bootes, Corona Borealis are circled
and shown in panels (a)--(e), respectively. The line corresponds to the
regression relation
$R_e \propto \sigma^{0.98\pm0.06} \langle I_e \rangle ^{-0.56\pm0.04}$.}
\end{figure*}

\begin{table*}
\setcaptionmargin{0mm} \onelinecaptionstrue
\captionstyle{flushleft}
\caption{Parameters of relations}
\medskip
\begin{tabular}{l|c|r|c|c}
\hline
\multicolumn{1}{c|}{Relation}              & $a$ & \multicolumn{1}{c|}{$b$}  &  Normalization & Scatter \\
\hline
$N=101$, $\sigma > 400$~km\,s$^{-1}$                &               &                &        &     \\
I~~~$L_K/L_{\odot}  = a\,\log R_e + b\,\log \sigma + c$    & $0.77\pm0.09$ & $1.41\pm0.12$  &  6.69  &0.11 \\
II~~$\log R_e ({\rm kpc}) = a \log \sigma - b \log \langle I_e \rangle  + c$  & $1.14\pm0.10$ & $-0.81\pm0.02$ &  4.78  &0.09 \\
III~$\log R_e ({\rm kpc}) = a\log \sigma - b\,\log \langle I_e \rangle  + c$ & $0.98\pm0.06$ & $-0.56\pm0.04$  & 3.57  &0.07 \\
\hline
$N=172$                                              &               &                &        &     \\
I~~~$L_K = a\,\log R_e + b\,\log \sigma + c$                & $0.64\pm0.08$ & $1.45\pm0.08$  &  6.95  &0.13 \\
II~~$\log R_e ({\rm kpc}) = a\,\log \sigma - b\,\log \langle I_e \rangle  + c$  & $1.06\pm0.06$ & $-0.73\pm0.01$ &  4.42  &0.09 \\
III~$\log R_e ({\rm kpc}) = a\,\log \sigma - b\,\log \langle I_e \rangle  + c$ & $0.90\pm0.04$ & $-0.48\pm0.03$  & 3.32  &0.07 \\
\hline
Leo Supercluster, $\langle z_h \rangle=0.034$, $N=10$           &               &                &        &    \\
I~~~$L_K = a\,\log R_e + b\,\log \sigma + c $              & $0.83\pm0.45$ & $1.20\pm0.44$  &  7.04  &0.08 \\
II~~$\log R_e ({\rm kpc}) = a\,\log \sigma - b\,\log \langle I_e \rangle  + c$  & $1.03\pm0.38$ & $-0.86\pm0.07$  &  5.35  &0.07 \\
III~$\log R_e ({\rm kpc}) = a\,\log \sigma - b\,\log \langle I_e \rangle  + c$ & $0.96\pm0.12$ & $-0.42\pm0.16$  & 2.80  &0.05 \\
\hline
Hercules Supercluster, $\langle z_h \rangle=0.031$, $N=9$       &               &                &        &     \\
I~~~$L_K = a\,\log R_e + b\,\log \sigma + c$               & $0.55\pm0.22$ & $1.53\pm0.27$  &  6.98  &0.08 \\
II~~$\log R_e ({\rm kpc}) = a\,\log \sigma - b\,\log \langle I_e \rangle + c$  & $1.05\pm0.18$ & $-0.69\pm0.04$  &  4.26  &0.07 \\
III~$\log R_e ({\rm kpc}) = a\,\log \sigma - b\,\log \langle I_e \rangle + c$ & $1.04\pm0.10$ & $-0.62\pm0.09$  & 3.80  &0.05 \\
\hline
Ursa Major Supercluster, $\langle z_h \rangle=0.055$, $N=11$    &               &                &        &     \\
I~~~$L_K = a\,\log R_e + b\,\log \sigma + c$               & $1.07\pm0.54$ & $1.53\pm0.40$  &  5.50  &0.12 \\
II~~$\log R_e ({\rm kpc}) = a\,\log \sigma - b\,\log \langle I_e \rangle  + c$   & $1.65\pm0.35$ & $-1.08\pm0.20$ &  5.07  &0.12 \\
III~$\log R_e ({\rm kpc}) = a\,\log \sigma - b\,\log \langle I_e \rangle + c$       & $0.71\pm0.28$ & $-0.30\pm0.17$ &  2.71  &0.14 \\
\hline
Bootes Supercluster, $\langle z_h \rangle=0.071$, $N=11$        &               &                &        &     \\
I~~~$L_K = a\,\log R_e + b\,\log \sigma + c$               & $0.88\pm0.15$ & $0.73\pm0.19$  &  8.22  &0.08 \\
II~~$\log R_e ({\rm kpc}) = a\,\log \sigma - b\,\log \langle I_e \rangle  + c$   & $0.65\pm0.17$ & $-0.89\pm0.04$ &  6.61  &0.07 \\
III~$\log R_e ({\rm kpc}) = a\,\log \sigma - b\,\log \langle I_e\rangle + c$        & $0.65\pm0.12$ & $-0.77\pm0.10$ &  5.74  &0.06 \\
\hline
Corona Borealis Supercluster, $\langle z_h \rangle=0.072$, $N=8$                         &               &                &        &     \\
I~~~$L_K = a\,\log R_e + b\,\log \sigma + c $              & $1.09\pm0.54$ & $1.43\pm0.30$  &  5.79  &0.10 \\
II~~$\log R_e ({\rm kpc}) = a\,\log \sigma - b\,\log \langle I_e\rangle + c$   & $1.55\pm0.32$ & $-1.10\pm0.09$ &  6.49  &0.11 \\
III~~$\log R_e ({\rm kpc}) = a\,\log \sigma - b\,\log \langle I_e\rangle + c$        & $0.88\pm0.30$ & $-0.43\pm0.25$ &  3.09  &0.06 \\
\hline
\multicolumn{5}{p{15cm}}{\footnotesize I---FP from equation (5); II---FP  from equation (6); III---FP from equation (7).}
\end{tabular}
\end{table*}
\renewcommand{\baselinestretch}{1}

In Fig.~4 this relation relative to the major axis $\log R_e$ is given for the entire sample, against which each galaxy supercluster is shown as large circles. Relation (7) has a smaller scatter and is suitable for measuring the relative distances of galaxy systems if $R_e$ is measured in arcseconds.

The FP we constructed (Fig.~3) is consistent, within errors, with the results of Schaeffer et al. (1993) and D'Onofrio et al. (2013), which used other methods for determining the effective radius  $R_e$ of galaxy systems: either by fitting the Vaucouleurs profile into the surface brightness (luminosity) profile, or using other profiles (for example, Nubble, King). It should be noted that superclusters of galaxies as a whole are not virialized structures, but consist of virialized groups and clusters of galaxies,---therefore we can talk about the FP of members of superclusters of galaxies. General relationships for the entire sample and separately for each supercluster of galaxies are given in Table~1, where Roman numerals show: I---FP from equation~(5); II---FP from equation~(6); III---FP from equation~(7). Table~1 also shows the FP coefficients for the full sample ($N=172$).

If we consider the FP of each supercluster in more detail (Fig.~3), we can note that some systems deviate from the general dependence for the supercluster. For example, the cluster A\,1831B (Fig.~3d) deviates from the general dependence for the Bootes supercluster. As we showed (see Kopylov and Kopylova, 2010), the cluster A\,1831B has a dynamic mass (velocity dispersion) apparently corresponding to such stages of evolution when the merging of smaller subcomponents has occurred almost strictly along the line of sight, but complete relaxation has not yet occurred.

Figure~5 presents the FP of groups and clusters of galaxies along with the FP of early-type galaxies. To do this, we converted the $L_K$-luminosity of galaxy systems taking into account the color $(r-K)\sim 3$ into $r$-filter luminosity (SDSS)---$L_r$. We took the rounded color value (($r_{\rm pet}-extinc)-K$) calculated for our sample: the average color of the systems varies in the interval [2.5--3.0]. Early-type galaxies were extracted from the SDSS (DR8) catalog when determining peculiar velocities in superclusters. A description of the extraction method and the results obtained for galaxy superclusters are published Kopylova and Kopylov (2014; 2017). The FP of 2111 early-type galaxies is determined as for galaxy clusters by regression with respect to $L_r$. It can be noted that the shape of the FP of both galaxies and galaxy systems agrees well with each other, as we previously obtained for 94~galaxy systems (Kopylova and Kopylov, 2016). The zero point of the FP of galaxy groups and clusters increases if we take into account the mass-to-luminosity ratio as the fourth parameter, as shown in the paper of Kopylova and Kopylov (2016).

\section{DETERMINATION OF DISTANCES AND PECULIAR VELOCITIES OF GALAXY SYSTEMS USING THE FUNDAMENTAL PLANE}

\begin{figure*}[]
\setcaptionmargin{3.5mm} \onelinecaptionstrue \captionstyle{normal} \vspace{-2mm}
\includegraphics[scale=0.55,angle=-90]{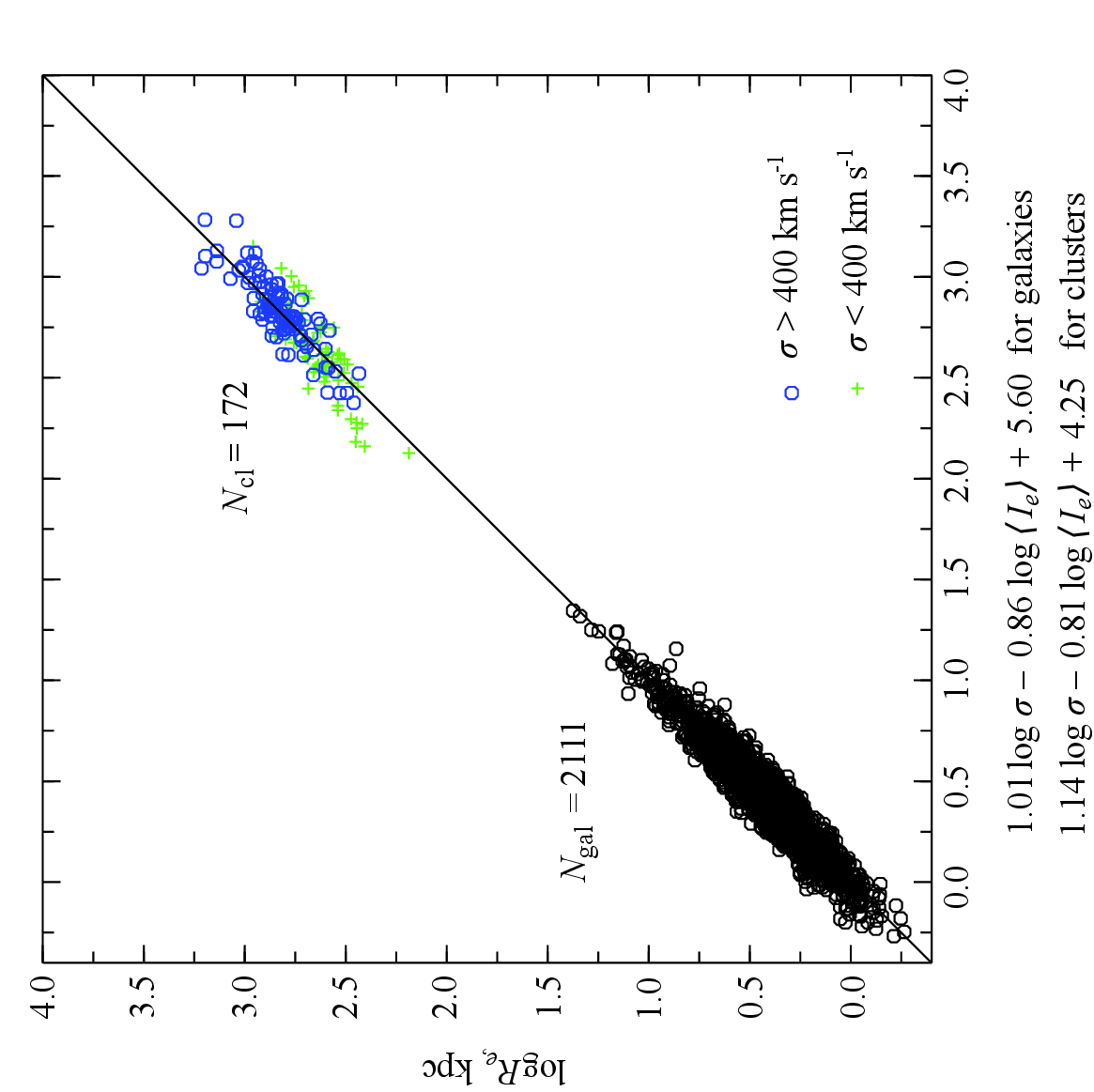}
\caption{Fundamental plane of galaxies and galaxy systems along the long axis, $\log R_e$. Blue empty circles show galaxy systems with $\sigma > 400$~km\,s$^{-1}$, green crosses---with $\sigma \leq 400$~km\,s$^{-1}$.}
\end{figure*}

If the radius $\log R_e$, the observed quantity, is measured in arcseconds, then the FP zero point 

\begin{equation}
\gamma = \log R_e-0.98(\pm0.06) \log \sigma+0.56(\pm0.04) \log \langle I_e \rangle
\end{equation}
varies with the distance of galaxy systems.

Figure~6 shows the obtained individual observed distances (zero points) of groups and clusters of galaxies. Crosses over empty circles mark galaxy clusters from the galaxy superclusters. The thick line shows the expected Hubble dependence between radial velocity and distance, calculated by us when constructing the FP of early-type galaxies (Kopylova and Kopylov, 2017) and shifted by the difference in the zero points. In this case, the comoving distance is

\begin{equation}
D\approx \cfrac{cz}{H_0}\Bigg(1- \cfrac{z(1-q_0)}{2}\Bigg)\approx \cfrac{cz}{H_0}\Big(1-0.225z\Big) \nonumber
\end{equation}
(Peebles approximation (Peebles, 1993)), angular size distance is

\begin{equation}
D\approx \cfrac{cz}{H_0}\Bigg(\cfrac{1- 0.225z}{1+z}\Bigg) \nonumber
\end{equation}

Since the surface brightness of galaxy clusters is determined by us from an area measured in square kiloparsecs, and not in square arcseconds as in galaxies, therefore the expected Hubble dependence is shifted by the value of the difference in the zero points. We also took into account the correction for the cosmological dimming of the surface brightness $\log \langle I_e \rangle $ of galaxy clusters $-4\,\log(1+z)$. The logarithmic $rms$ scatter $\gamma$ is equal to 0.07 ($N$ = 172, Table~1),  which corresponds to a 16\% error in determining the distance to a single cluster. Deviations from the Hubble dependence (Fig.~6) also have a scatter of 0.07 for galaxy clusters with \mbox{$\sigma$ > 400~km\,s$^{-1}$}, for groups of galaxies~--- 0.117. Based on galaxy clusters, we can determine the deviation of the Hubble constant we accepted to be 16\% (about $ \pm11$~km\,s$^{-1}$Mpc$^{-1}$).

\begin{figure*}[]
\setcaptionmargin{3.5mm} \onelinecaptionstrue \captionstyle{normal}
\includegraphics[scale=0.55,angle=0]{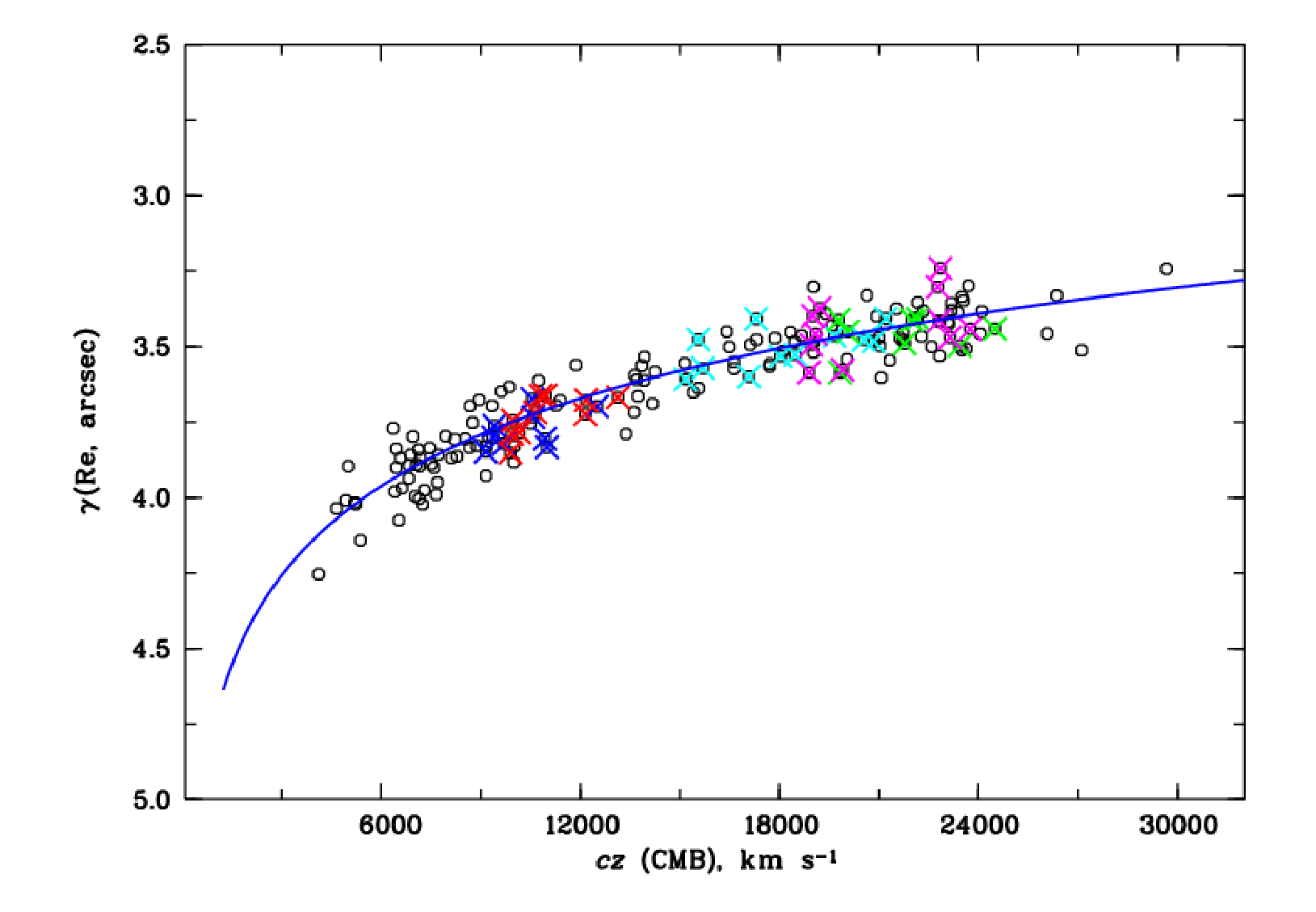}
\caption{Dependence of individual distances in arcseconds, zero points of the fundamental plane $\gamma$, on the radial velocity of galaxy systems $cz$ (CMB). Empty circles correspond to 172 galaxy systems. Green crosses indicate members of the CrB supercluster, magenta and cyan---members  of the Bootes and UMa superclusters, blue crosses indicate members of the  Hercules supercluster, red---members of the Leo supercluster. The thick blue curved line corresponds to the Hubble relationship between radial velocity and distance for a model with the cosmological parameters: $\Omega_m=0.3$, $\Omega_{\Lambda}=0.7$, $H_0=70$~km\,s$^{-1}$ Mpc$^{-1}$.}
\end{figure*}

\begin{figure*}[]
\setcaptionmargin{3.5mm} \onelinecaptionstrue \captionstyle{normal} \vspace{-2mm}
\includegraphics[scale=0.55,angle=0]{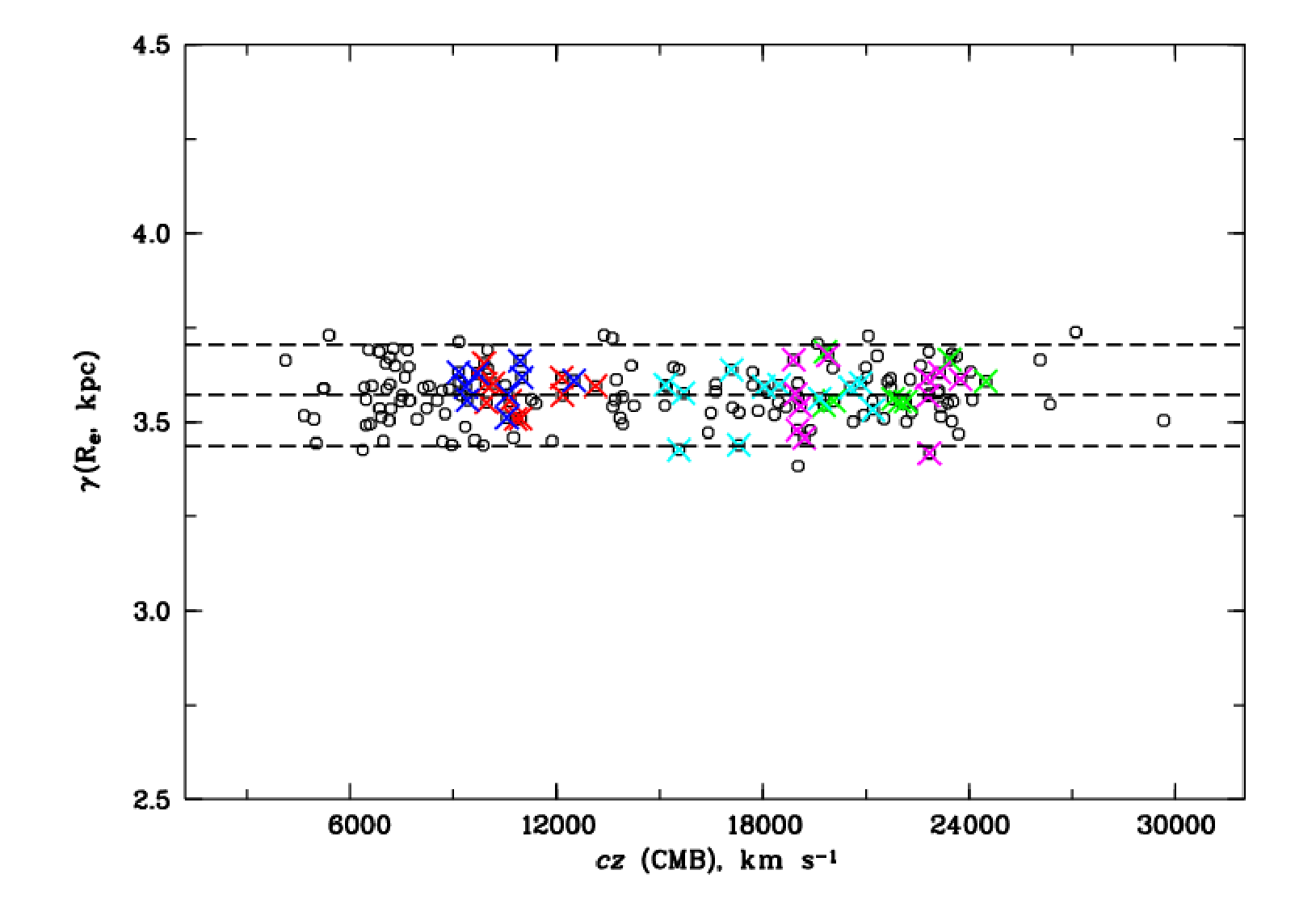}
\caption{Dependence of individual distances of galaxy systems in kpc in the comoving  system, for the model we adopted, on the radial velocity of galaxy systems $cz$ (CMB). Empty circles correspond to 172~galaxy systems. The members of the CrB, Bootes and UMa superclusters are marked with green, magenta and cyan crosses, and the members of the Hercules and Leo supercluster are marked with blue and red crosses.}
\end{figure*}

Figure~7 shows the same systems of galaxies in the comoving coordinate system, where the relative distances $\log R_e$ are converted into kiloparsecs according to the model we adopted. In practice, the difference between the cluster zero point $\gamma_{\rm cl}$ and the overall mean zero point $\langle \gamma \rangle$ is used to calculate photometric redshifts 

\begin{equation}
z_{\rm FP}=z_{\rm CMB}\times10^{(\langle \gamma \rangle -\gamma_{\rm cl})}.
\end{equation}

Then the peculiar velocities of galaxy clusters in the comoving system are equal to the difference between the spectroscopic and photometric redshifts:

\begin{equation}
V_{\rm pec}=c\,(z_{\rm CMB}-z_{\rm FP})/(1+z_{\rm FP}).
\end{equation}

We have obtained the following peculiar velocities of galaxy superclusters CrB ($N = 8$), Bootes \mbox{($N = 11$)}, UMa ($N = 11$), Her ($N = 9$) and Leo ($N = 10$): \mbox{$V_{\rm pec}=+740\pm890$~km\,s$^{-1}$}, $-340\pm1200$~km\,s$^{-1}$, \linebreak 
$-600\pm830$~km\,s$^{-1}$, $+530\pm350$~km\,s$^{-1}$, $+44\pm360$~km\,s$^{-1}$  respectively. Average peculiar velocity of galaxy superclusters relative to the CMB is \mbox{$V_{\rm pec} = +75\pm360$~km\,s$^{-1}$}. We can conclude that the peculiar motions of such giant structures are insignificant and practically do not exceed measurement errors.

\section{CONCLUSIONS}

Our sample, consisting of 172 galaxy systems, has redshifts of $0.012 < z < 0.090$, of which 71~groups of galaxies have radial velocity dispersions of \linebreak \mbox{$\sigma \leq 400$~km\,s$^{-1}$}, the remaining 101~systems are clusters of galaxies with \mbox{$\sigma > 400$~km\,s$^{-1}$}. In this work, we observationally determined the splashback radius (dark halo boundary) of clusters and groups of galaxies (radius $R_{\rm sp}$)  and measured the effective radius $R_e$, within which $N/2$ galaxies are observed, from the observed integrated distribution of the number of galaxies as a function of the squared distance from the center (Fig.~1c). The work also shows that the relationships between the characteristics of galaxy systems (fundamental planes) obtained in this way are consistent with the results of other authors using other methods (Schaeffer et al., 1993; Adami et al., 1998; D'Onofrio et al., 2013).

The main results of this study are given below:

1. The general fundamental plane of groups and clusters of galaxies in the $K$-band has the form: \mbox{$L_K = R_e^{0.64\pm0.08} \sigma^{1.45\pm0.06}$} with $rms = 0.13$. If we take only galaxy systems with $\sigma$ > 400~km\,s$^{-1}$, we obtain $L_K = R_e^{0.77\pm0.09} \sigma^{1.41\pm0.12}$ with a smaller $rms = 0.11$.

2. For the first time, the fundamental planes of members of 5 superclusters of galaxies have been constructed: Leo, Hercules, Ursa Major, Bootes, Corona Borealis, the shapes of which, within the limits of error, are consistent with each other and with the general FP of 172 galaxy systems. The FP zero points of cluster systems are different, but this fact does not affect the determined peculiar velocities due to their large errors.

3. For the first time, the relative distances of groups/clusters of galaxies along their fundamental plane have been determined. For the superclusters of galaxies Corona Borealis, Bootes, Ursa Major, Hercules and Leo, we obtained the following estimates of peculiar velocities: $V_{\rm pec} = +740\pm890$~km\,s$^{-1}$, $-340\pm950$~km\,s$^{-1}$, $-600\pm830$~km\,s$^{-1}$, $+530\pm350$~km\,s$^{-1}$, $+44\pm360$~km\,s$^{-1}$. The average peculiar velocity of galaxy superclusters relative to the CMB is  +$75\pm360$~km\,s$^{-1}$. It can be concluded that the peculiar motions of the superclusters considered are insignificant and practically do not exceed measurement errors.

\begin{acknowledgments}
This research has made use of the NASA/IPAC Extragalactic Database (NED, \url{http://nedwww.} \linebreak \url{.ipac.caltech.edu}), which is operated by the Jet Propulsion Laboratory, California Institute of Technology, under contract with the National Aeronautics and Space Administration, Sloan Digital Sky Survey (SDSS, \url{http://www.sdss.org}), which is supported by Alfred P. Sloan Foundation, the participant institutes of the SDSS collaboration, National Science Foundation, and the United States Department of Energy and Two Micron All Sky Survey (2MASS, \url{http://www.ipac.}\linebreak\url{.caltech.edu/2mass/releases/allsky/}).
\end{acknowledgments}


\section*{CONFLICT OF INTEREST}
The authors of this work declare that they have no conflict of interest.

\begin{center}
\refname
\end{center}


\end{document}